\documentclass[doublecol]{epl2}
\usepackage{amsmath,amssymb,bm}
\usepackage{graphicx,color}

\newcommand{\hide}[1]{{}}
\newcommand{\delete}[1]{\textcolor{red}{}}

\title{Correlations of strain and plasticity in flowing foam}

\author{V. Chikkadi\inst{1,2} \and E. Woldhuis\inst{2}, M. van Hecke\inst{3,4} \and P. Schall\inst{1}}
\shortauthor{V. Chikkadi \etal}

\institute{
  \inst{1} Institute of Physics, University of Amsterdam, Science Park 904, 1098 XH Amsterdam, The Netherlands,\\
  \inst{2} Instituut-Lorentz, Universiteit Leiden, Postbus 9506, 2300 RA Leiden, The Netherlands, \\
  \inst{3} Huygens-Kamerlingh Onnes Lab, Universiteit Leiden, Postbus 9504, 2300 RA Leiden, The Netherlands\\
  \inst{4} Amolf, Science Park 104, 1098 XG Amsterdam, The Netherlands.
}


\abstract{Via simulations of flowing foam, we connect the high and intermediate density regimes of complex fluid flows into a consistent microscopic picture of deformation. While at and above the jamming transition, elastic correlations lead to strong spatial organization of the flow field, below jamming, the slowly diminishing elastic correlation length leads to slowly ceasing spatial organization, which is nevertheless still present down to densities far below jamming.
We show that the long-range correlated flow field arises from the superposition of quadrupolar strain fields of shear zones with highly correlated positions, strengths and orientation. These interactions are still pertinent below jamming, where they systematically weaken with the slowly diminishing elastic correlation length. These results demonstrate the ubiquity and importance of elastic correlations in the flow of complex fluids even below the jamming transition, and motivate a scale-bridging description of their flow over wide ranges of density from solid to fluid.
}

\begin{document}

\maketitle

Disordered packings of foams, emulsions, colloidal suspensions and granular particles all exhibit a rigidity transition to jammed solids when packed densely \cite{Cates1998, LiuNagel1998,vanHeckeReview,OHern2002,OHern2003,vanSaarloos2006}. This rigidity vanishes at the jamming transition, where the packing approaches its stability limit: the average number of contacts with neighboring particles approaches a critical stability limit, and the shear moduli vanish with well-known scaling relations\cite{OHern2002, OHern2003}. Concomitantly with the loss of elasticity, non-affine fluctuations become increasingly important, and floppy modes indicate the increasing susceptibility of the material to applied stress~\cite{vanSaarloos2006,OlssonTeitel,Xu2010,Manning2011,vanHecke}. At unjamming the elastic moduli of the static packing vanish and the material is irreversibly affected by the smallest applied force\cite{OHern2002, OHern2003,vanHecke}.

The situation changes qualitatively when the material is subjected to flow. In steady-state flow, particle contacts are constantly broken and reformed, leading to a dynamic scenario of continuously changing, transient particle contacts~\cite{Heussinger09}. At the same time, the structural rearrangements have to be relaxed to the boundaries by some long-range displacement field. The nature of these fields as the density decreases near to and below the static jamming transition remains unclear. In the deeply jammed state, the material exhibits pronounced elasticity that causes correlations in the flow, and provides the long-range field that transfers the local relaxation. In this regime, simulations as well as experiments have established that the flow of dense foams, emulsions and suspensions is governed by local shear transformation zones~\cite{Argon79,Falk98,Schall07} surrounded by a long-range quadrupolar elastic strain field~\cite{Eshelby}. While this long-range elastic field - an essential feature of elastic materials - provides interactions between transformation zones and leads to strongly correlated flow, it is unclear what happens near jamming, and how the flow ultimately crosses over to the Newtonian regime far below jamming.
Despite its central importance for many applications of complex fluids in industry and consumer products, this intermediate flow regime remains poorly understood.

In this letter we elucidate just this intermediate flow regime between the strongly correlated high-density, and the low-density Newtonian regime using simulations of flowing foam. We show that while similar to static packings, the flow field becomes increasingly delocalized with decreasing density, it is still uniquely determined by elastic correlations at and below the jamming density, where the elasticity of the static packing vanishes. The robust power-law correlations persist across the jamming transition and become systematically truncated by a slowly diminishing cut-off below jamming. We demonstrate that the non-trivial power-law exponent arises from the superposition of correlated quadrupolar strain fields that due to ceasing elastic interactions become increasingly delocalized. These results demonstrate the abundance of robust elastic-like correlations even in regimes where the elasticity of the static packing vanishes, highlighting the crucial role of elastic correlations in complex fluid flows, and motivating a universal scale-bridging framework over a wide range of density.

\begin{figure}
\centering
\includegraphics[height=0.2\textwidth]{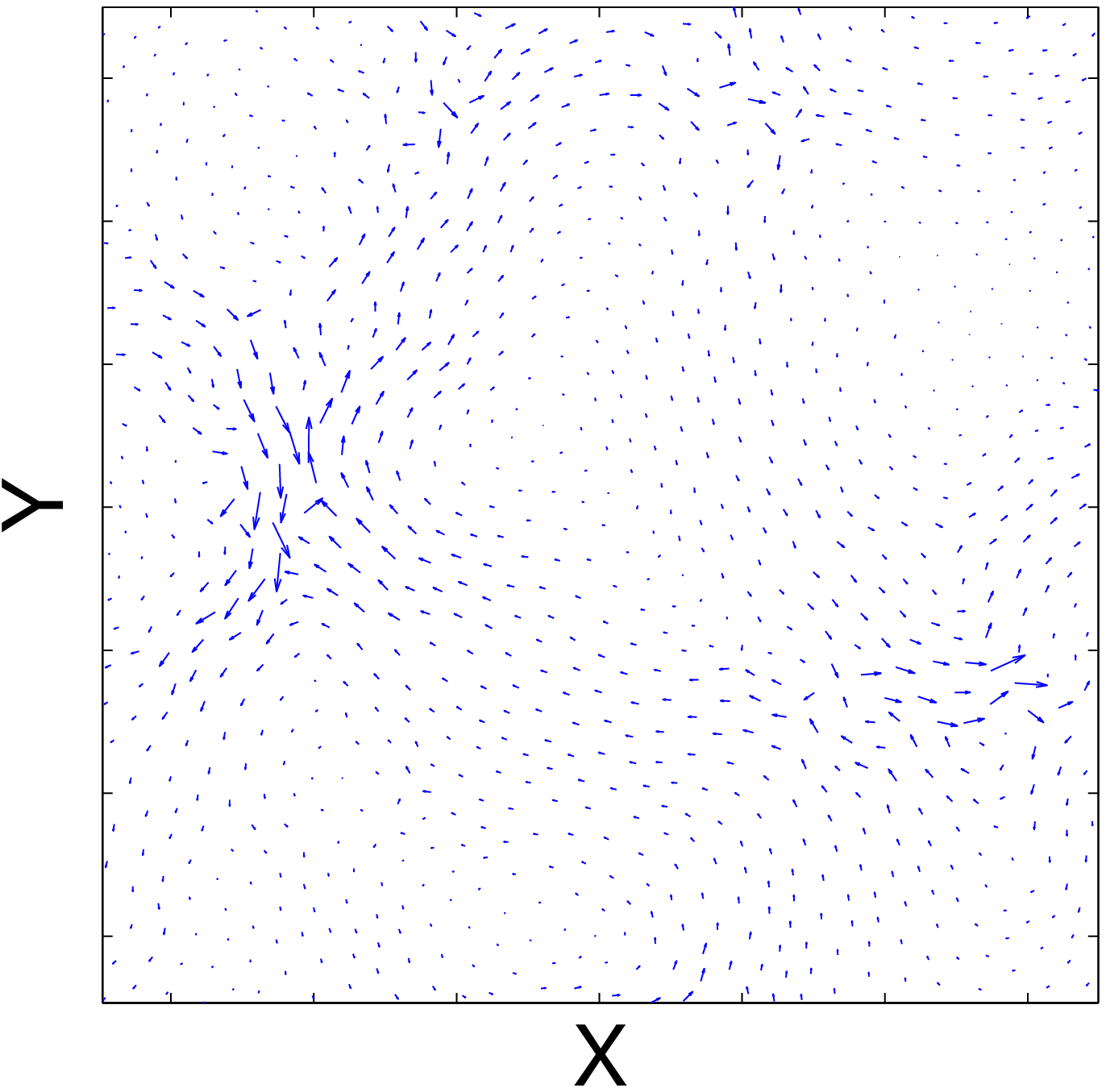}
\includegraphics[height=0.2\textwidth]{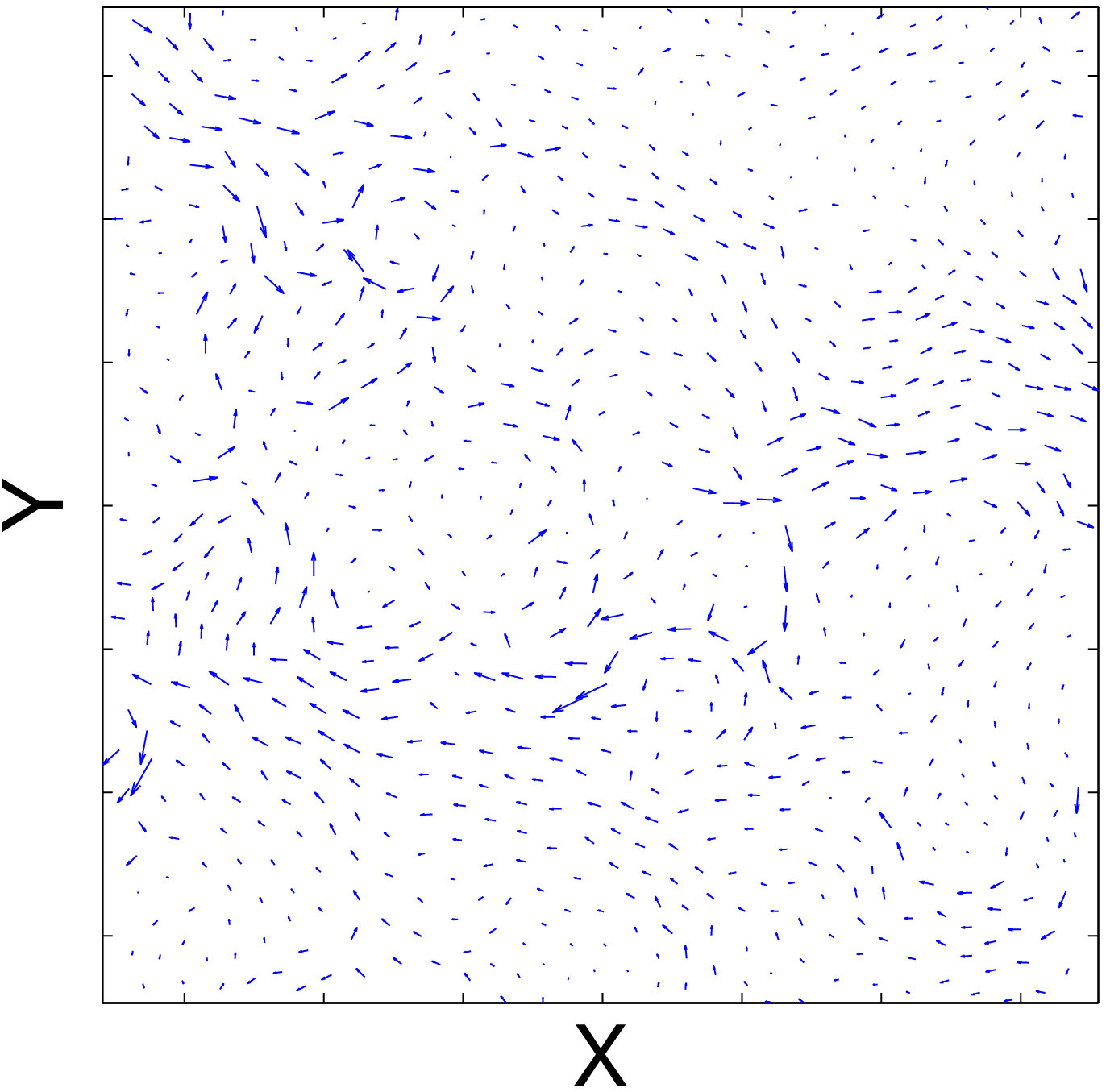}
\put(-227,-9){(a)}
\put(-111,-9){(b)}
\quad
\caption{Non-affine displacements during the flow of foam at shear-rate $\dot{\gamma}=10^{-5}$ for bubble volume fractions of $\phi=1.0$ (a) and $\phi \sim \phi_J = 0.8424$ (b). The flow field becomes delocalised.}
\label{fig1}
\end{figure}

{\em Model:} We probe correlations of flowing foam in Durian's two-dimensional bubble model~\cite{Durian1995},
which captures many aspects of complex fluid flows above, near, and below jamming~\cite{vanHeckeReview,OlssonTeitel,Tigh}. In this model, particles or bubbles
are represented by disks which interact through purely repulsive elastic and viscous contact forces. Inertia is absent, and
the sum of elastic and dissipative forces on each particle balance at all times.
Elastic forces are proportional to the disk overlap $f^{el}_{ij}=k(R_i+R_j-r_{ij})$, where $r_{ij}:=|\bar{r}_j - \bar{r}_i|$ is the distance between bubble centers and $R_i$ is the radius of disk $i$. Viscous forces oppose the bubbles' relative velocity $\Delta \bar{v}_{ij} := \bar{v}_j- \bar{v}_i$ with magnitude $f^{visc}_{ij}=b|\Delta v_{ij}|$. The packing fraction $\phi$ controls the density, with the jamming density of order $\phi_J = 0.8424$ \cite{woldhuis2015}.
The strain rate $\dot{\gamma}$ is imposed via Lees-Edwards boundary conditions, which lead to a linear flow profile where $\langle v(y) \rangle = \dot{\gamma} ~ y ~ \mathbf{e}_{x}$, with $y$ and $x$ the transversal and stream-wise coordinates.
The unit cell contains a $50:50$ bidisperse mixture of $N = 1020 - 1210$ bubbles with size ratio $1.4:1$ to avoid crystallization. In the following we use particle diameter, elastic ($k$) and viscous ($b$) prefactors  to nondimensionalize our results. The global shear stress is obtained according to $\sigma_{xy}=\sigma_{tot}=1/(2V)\Sigma_{<ij>}r_{ij,x}(f^{el}_{ij,y}+f^{visc}_{ij,y})$, where $V$ is the area of the unit cell and the sum runs over contacting pairs. We focus here on slow shear, namely foam sheared at a rate of $\dot{\gamma}=10^{-5}$. Averages are taken over runs of total time $20/\dot{\gamma}$ after discarding the transient.

To probe fluctuations in the local strain field, we follow~\cite{Falk98,Schall07} and  determine, for each particle, the local strain from the displacement of a particle with respect to its nearest neighbors. We identify nearest neighbors as those separated by less than $r_{l}$, the radius of the larger bubbles. We subsequently determine the best affine deformation tensor ${\bf \Gamma}$ that transforms the nearest neighbor vectors, ${\bf d_i}$, over the applied strain interval $\delta \gamma$ \cite{Falk98}, by minimizing $D^2_{min} = (1/n) {\sum_{i=1}^{n}}({\bf d_i}(\gamma + \delta \gamma) - {\bf \Gamma}{\bf d_i}(\gamma))^2$. The symmetric part of ${\bf \Gamma}$ is the local strain tensor, whose off-diagonal component is the shear strain $\epsilon \equiv \epsilon_{xy}$, which we will use in the further analysis below.

\begin{figure}
\centering
\includegraphics[width=0.49\columnwidth,clip,viewport=225 0 740 450]{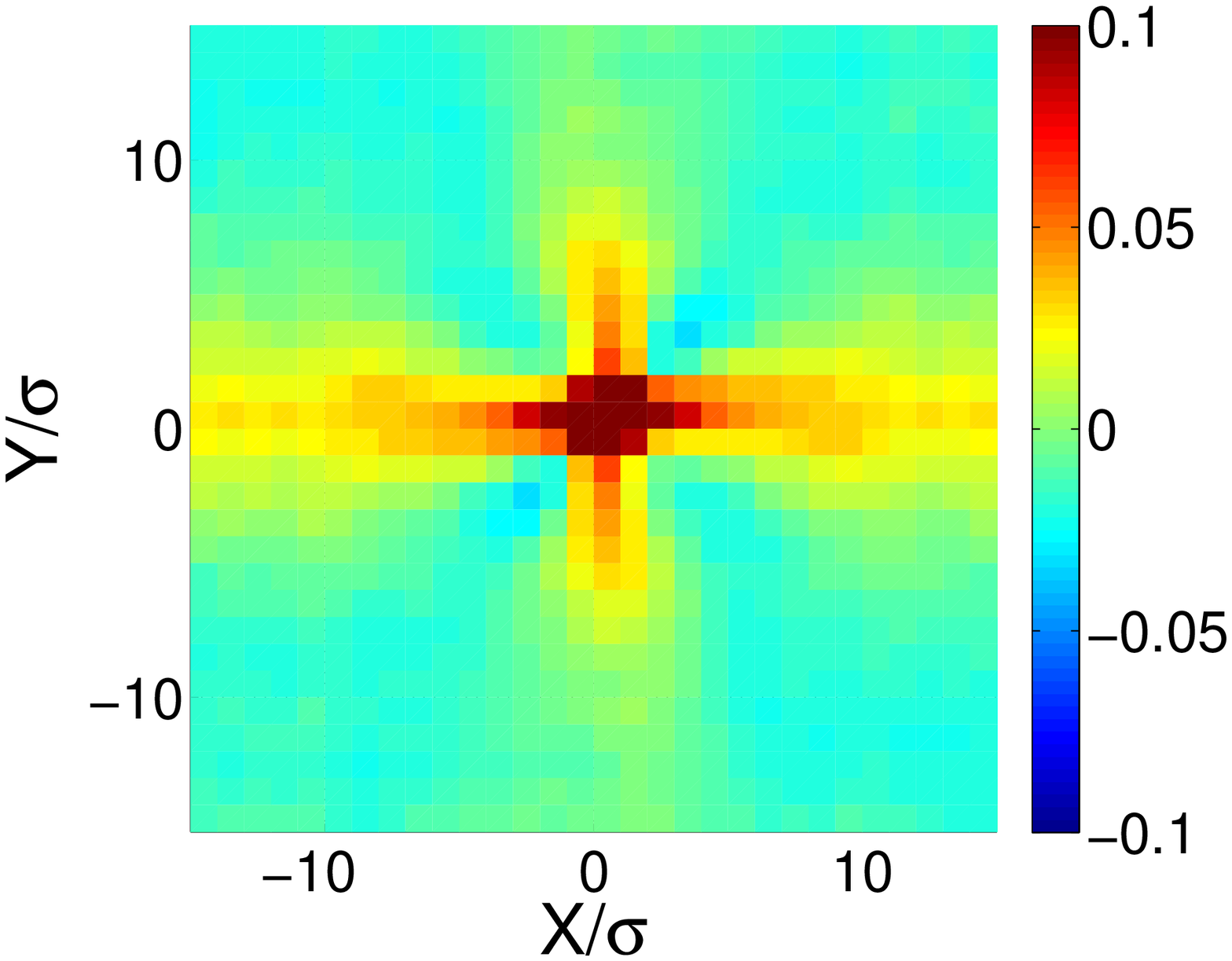}
\includegraphics[width=0.495\columnwidth,clip,viewport=233 0 733 400]{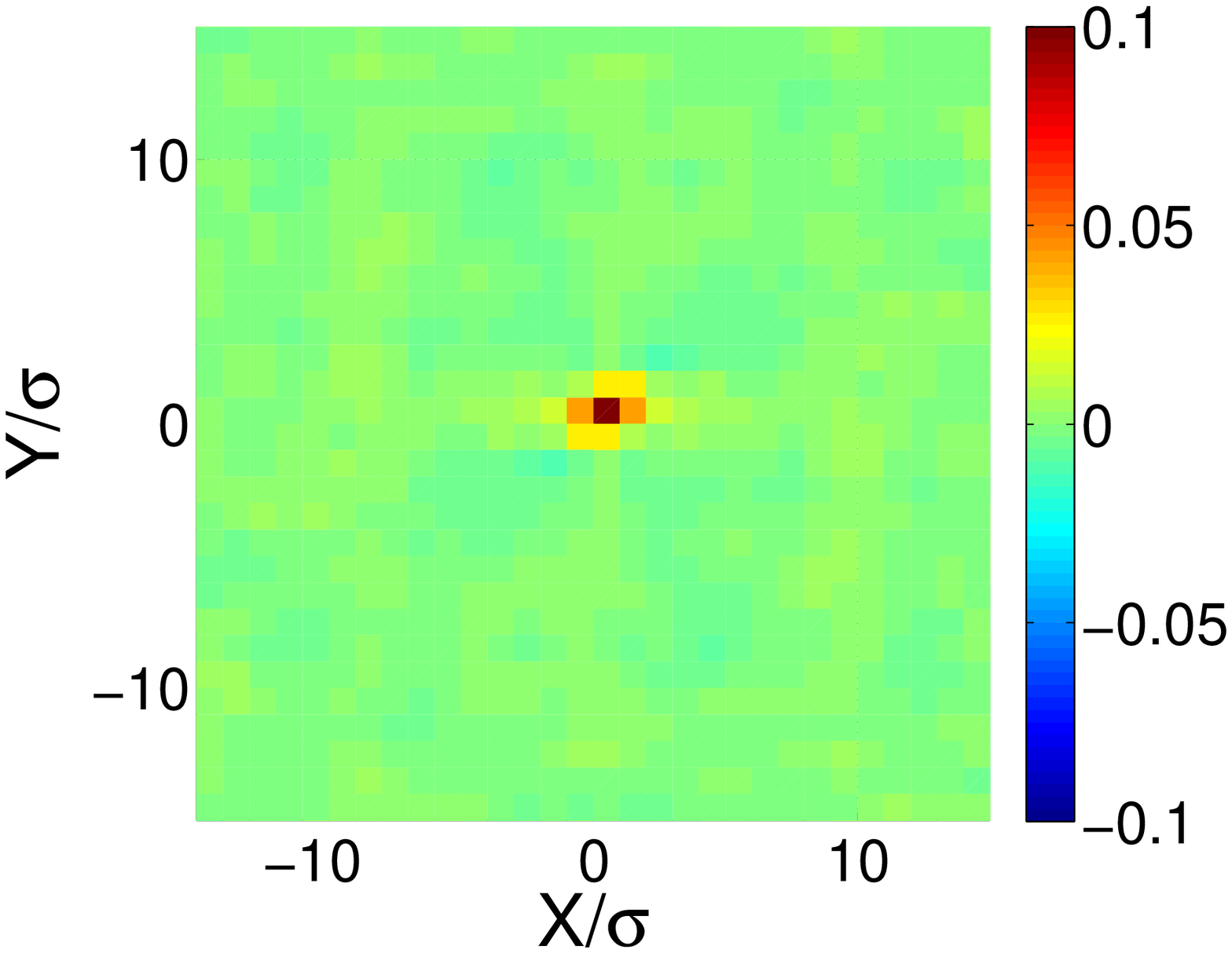}
\includegraphics[width=0.525\columnwidth,clip,viewport=231 0 740 450]{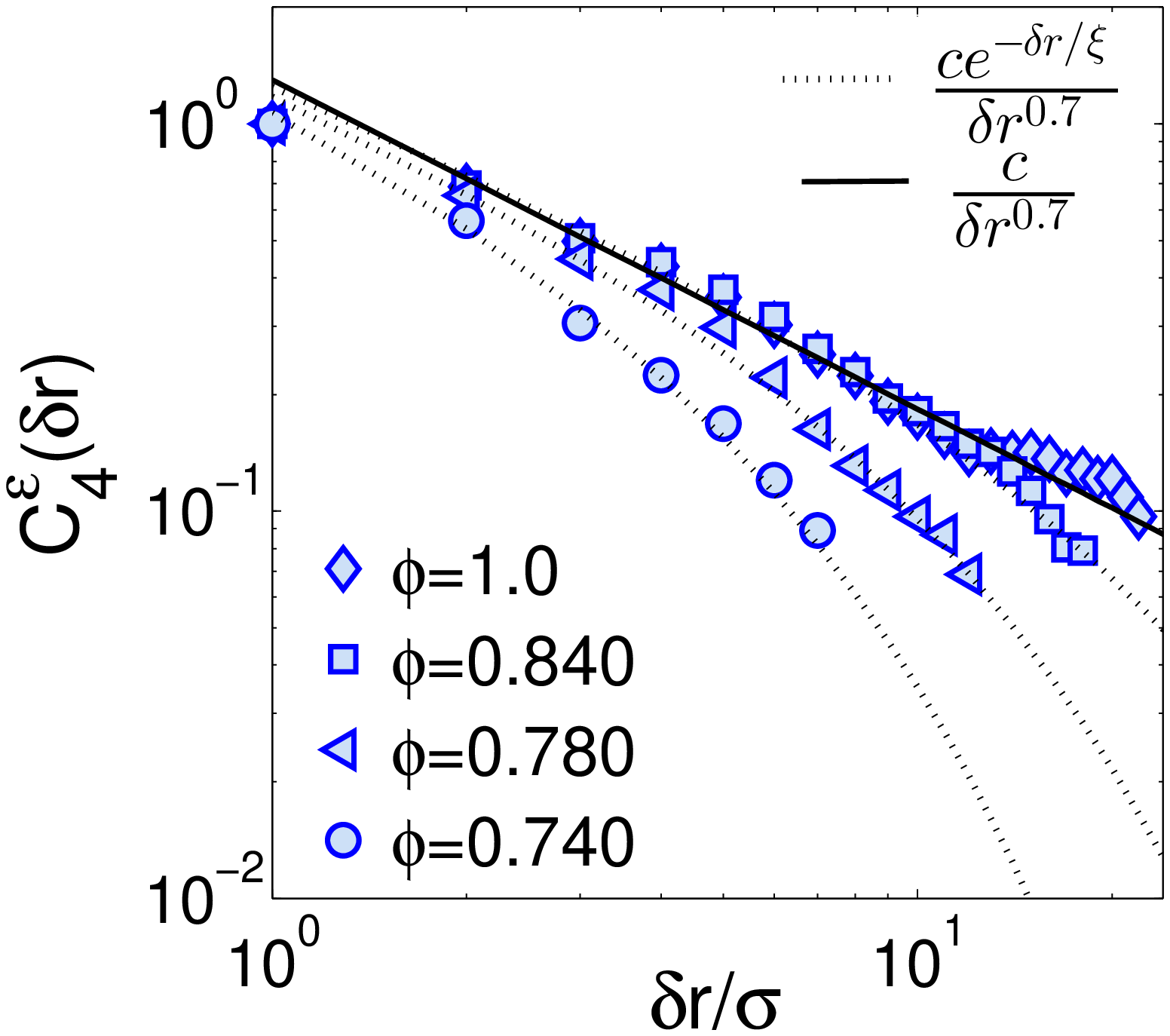}
\includegraphics[width=0.46\columnwidth,clip,viewport=15 0 600 550]{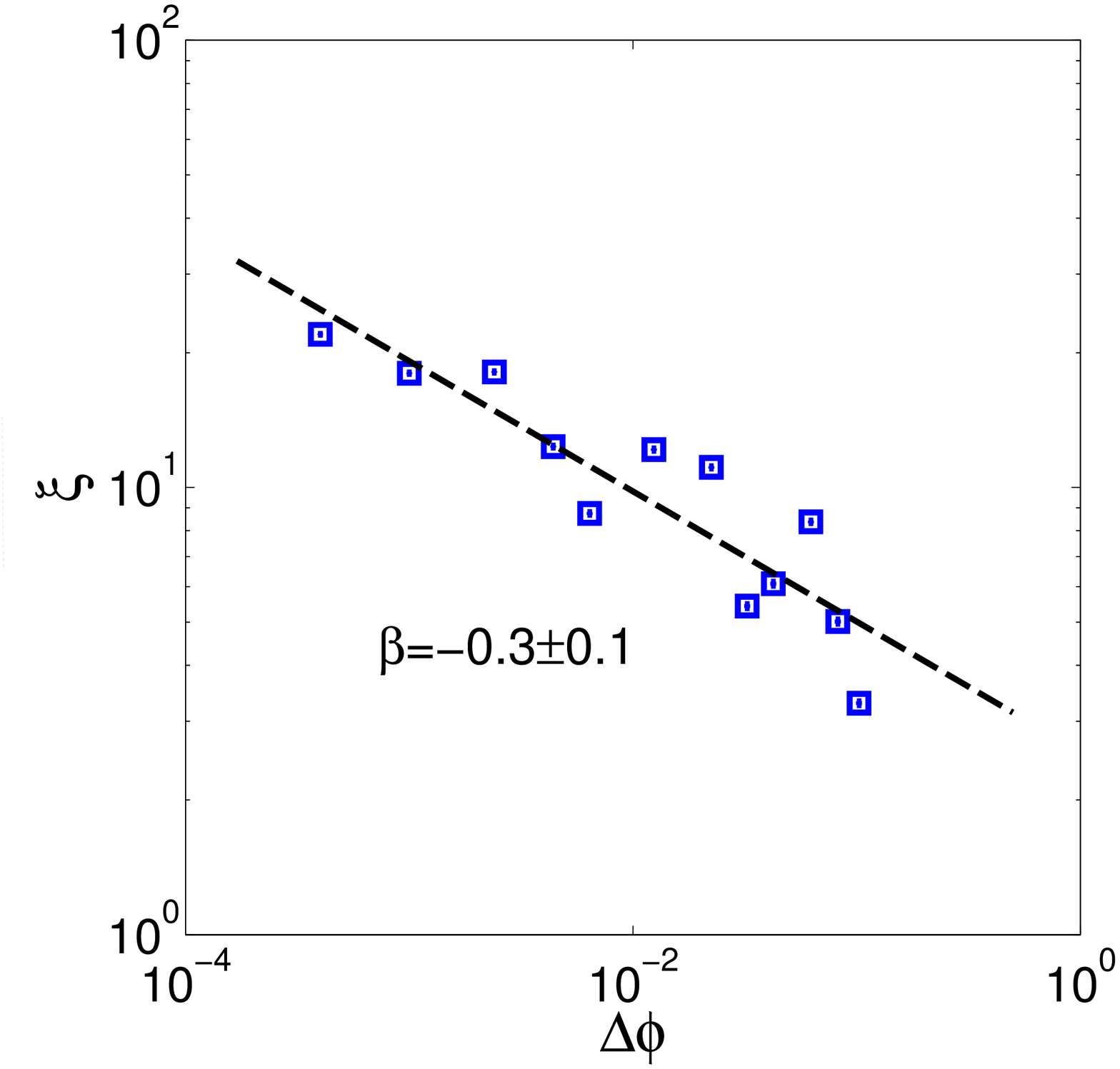}
\put(-240,136){(a)}
\put(-120,136){(b)}
\put(-240,-5){(c)}
\put(-120,-5){(d)}
\caption{Strain correlations and scaling. (a,b) Spatial correlations of strain for flowing foam with $\phi=1.0$ (a) and $\phi=0.740$ (b). Color indicates value of the normalized correlation function, see color bar. Symmetry change indicates loss of elasticity. (c,d) Decay of strain correlations and correlation length. (c) Radial decay of strain correlations projected onto the 4th circular harmonic. Correlations are power-law for $\phi>\phi_J$, and become increasingly short-ranged below $\phi_J$ (d) Correlation length ($\xi$) extracted from the decay of correlations in (c), as a function of distance $\Delta \phi$ from jamming. Error bars are twice the symbol size.}
\label{fig2}
\end{figure}

{\em Phenomenology:} Snapshots of the deformation field with mean flow subtracted reveal significant spatial and temporal heterogeneities, and moreover, a distinct trend in their qualitative nature with packing density $\phi$, as shown in Fig~\ref{fig1}. Well above the jamming density, the deformation field appears strongly localized with clear quadrupolar structures (Fig.~\ref{fig1}a), while at densities approaching $\phi_J$, rearrangements become more extended and the displacement field becomes delocalized (Fig. \ref{fig1}b). Both the strong spatial fluctuations and trends with packing density resemble those of quasistatic deformations, as studied extensively in the context of jamming \cite{vanHeckeReview,vanSaarloos2006}.

{\em Correlations:}
The question we wish to answer is, what do these fluctuations tell us about the underlying nature of the deformation field? In particular, how does the
high-density physical picture of local rearrangements coupled through an elastic background change when we approach the jamming point, or even go to densities below jamming?

To probe the underlying spatial organization of the instantaneous deformation fields, we focus on the local shear strain $\epsilon$, and compute spatial correlations according to ~\cite{chikkadi_schall11}
$C_\epsilon(\delta \bar{r}) = (\left< \epsilon(\bar{r} + \delta \bar{r}) \epsilon(\bar{r})\right> - \left< \epsilon(\bar{r})\right> ^{2}) / \sigma^2$,
where the squared standard deviation $\sigma^2 = \left< \epsilon(\bar{r})^{2} \right> - \left< \epsilon(\bar{r})\right>^{2}$.
As shown in Fig.~\ref{fig2}a and b, these correlations reveal the typical symmetries and range of strain fluctuations. For large densities (Fig.~\ref{fig2}a), $C_{\varepsilon}(\delta x,\delta y)$ has a distinct
four-fold symmetry stemming from the abundance of local quadrupolar strain fields that reflect the response of an elastic matrix to local shear transformations~\cite{chikkadi_schall11}. Surprisingly and as we will show in detail below, this quadrupolar elastic response remains essentially unchanged down to the jamming transition, where the instantaneous deformation field does not any more show clear quadrupoles (Fig.~1b). In fact, the quadrupolar symmetry of $C^{\varepsilon}(\delta x,\delta y)$  vanishes only far below jamming, where the correlation function eventually becomes isotropic and short ranged (Fig.~\ref{fig2}b). The observation that elastic correlations behind this flow field remain important near jamming is also consistent with observations of system-spanning velocity correlations of the instantaneous flow field \cite{Heussinger09,woldhuis2015}.

To quantify the gradual loss of elasticity, we investigate the range of quadrupolar correlations. We project each correlation function onto the corresponding circular harmonic, using $C_{4}^{\epsilon}(\delta r) = \int^{2\pi}_{0}C_{\epsilon}(\delta r, \theta) cos(4\theta) d\theta$, and study the radial decay of $C_{4}$. In  Fig.~\ref{fig2}(c) we show examples of $C_{4}(\delta r)$ for densities above, near and below jamming.
These show that down to $\phi_J$, the quadrupolar field is long ranged: strain correlations remain robust upon approaching the jamming point. Below jamming, the range of correlations decreases slowly until at $\phi \sim 0.74$, the correlation length has decreased down to a few particle diameters, and the correlation function appears closely isotropic. We extract a correlation length quantitatively by fitting the decay of correlations to $C_4^\epsilon \sim \delta r^{\alpha}~\exp(- \delta r/\xi)$ with the exponent $\alpha = - 0.7$, yielding excellent overlap with the measured correlation function. In  Fig.~\ref{fig2}(d) we plot the resultant correlation length $\xi$ as a function of density difference $\Delta \phi = \phi - \phi_J$, focussing on $\Delta \phi <0$, i.e., below jamming. We find that this length scale appears to diverge as $\xi \sim |\Delta \phi|^{-\beta}$, with the best fit for
$\beta = 0.3\pm0.1$.

\begin{figure}
\centering
\includegraphics[width=0.46\columnwidth,clip,viewport=270 0 750 450]{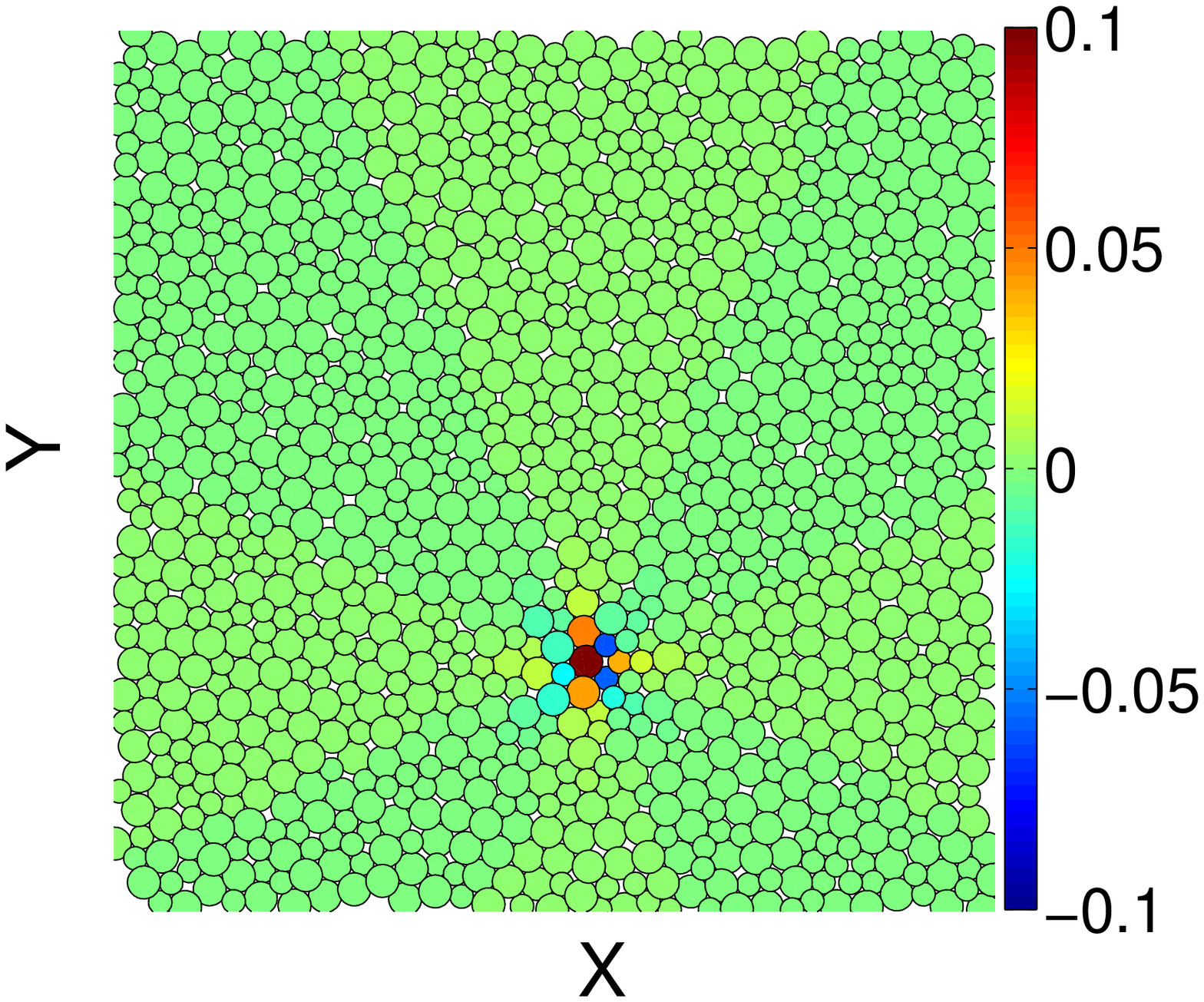}
\includegraphics[width=0.5\columnwidth,clip,viewport=230 0 740 450]{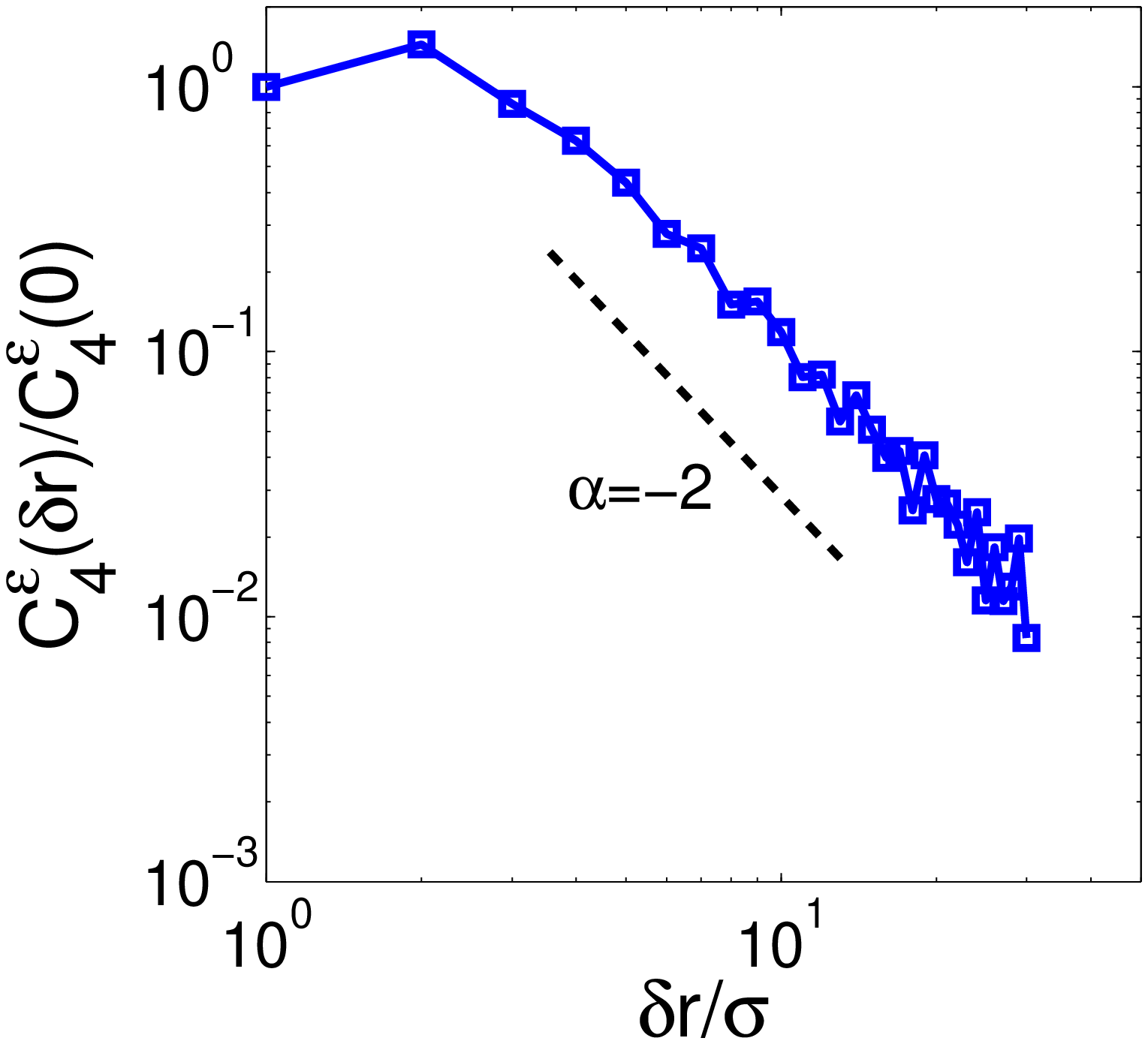}
\put(-235,-3){(a)}
\put(-115,-3){(b)}
\caption{(a) Synthetic strain field constructed for a single Eshelby inclusion. We selected a particle with large shear strain ($\epsilon=0.5$) as center of the inclusion, and computed the shear strain at the positions of all other particles using eq.~\ref{exy}.
(b) Decay of the projected strain correlations computed for the single Eshelby inclusion shown in (a).}
\label{fig3}
\end{figure}

{\em Quadrupoles:} The powerlaw radial decay of $C_4$ evidences strong organization of the flow in the jammed and near jammed regime, irrespective of $\Delta \phi$. What are the basic ingredients that generate this flow field?
The quadrupolar correlation function in Fig.~\ref{fig2}a suggests that the strain field is composed of an abundance of local quadrupolar strain fields, each of which is the elastic response to a local shear rearrangement. As we will show, the connection between quadrupoles and $C_4$ is subtle but firm. First, we note that a single Eshelby quadrupolar inclusion
decays as $\delta r^{-2}$, in contrast to $C_4$ which decays as $\delta r^{\alpha }$ with $\alpha \approx -0.7 > -2$. Second, we will show that realistic deformation fields are composed of a superposition of many of such quadrupoles, and that a superposition of quadropoles, precisely placed at the locations obtained in the simulations, reproduces a  powerlaw radial decay of $C_4$ with an exponent $\approx -1 > -2$. Third, a superposition of the same quadrupoles placed at random positions
does lead to $C_4 \sim 1/\delta r^2$, a similar decay as for the single inclusion. Taken together, this shows that a nontrivial spatial organization of quadrupoles is a necessary and also sufficient ingredient to reconstruct the nontrivial  powerlaw radial decay of $C_4$.

We start from a single quadrupole.  Since (weakly) jammed materials behave on average as a linear elastic matrix \cite{Ellenbroek2009}, we use the elastic shear strain produced by a single sheared inclusion of size $a$, with shear strain $\epsilon_0$,
\begin{equation}
\epsilon^{E}(\delta r, \theta) = \frac{\epsilon_0}{\pi}\frac{cos(4(\theta-\theta_0))}{(\delta r/a)^2} \quad \delta r>>a,
\label{exy}
\end{equation}
where $\theta_0$ denotes the orientation of the quadrupole ~\cite{Eshelby,Lemaitre}. Unless noted otherwise, we take $\theta_0=0$. Clearly, $\epsilon^{E}$ decays as $\delta r^{-2}$, which is qualitatively different from the decay of the correlation functions --- so a single Eshelby quadrupole cannot capture the radial decay of $C_4$.
Before studying superpositions of $\epsilon^{E}$, we show in Fig.~3 an example of a single quadrupolar field constructed by applying Eq.~(\ref{exy}) to a given packing. Strictly speaking, Eq.~(\ref{exy}) is only valid at sufficient distance from the center of the sheared inclusion, but here we use this form all the way to a central particle
with large shear strain ($\epsilon=0.5$), and compute the shear strain at the positions of all other particles using Eq.~(\ref{exy}). For the center particle, we take the actual shear strain of the particle as determined from the affine fitting; repeating this procedure for $\sim 1000$ realizations (separated by strains of $\sim 5 \cdot 10^{-4}$) we obtain the corresponding correlation function calculated for single, isolated shear events, which clearly decays as $1/\delta r^2 $ (Fig.~\ref{fig3}b).

\begin{figure}
\centering
\includegraphics[width=0.52\columnwidth,clip,viewport=150 0 757 450]{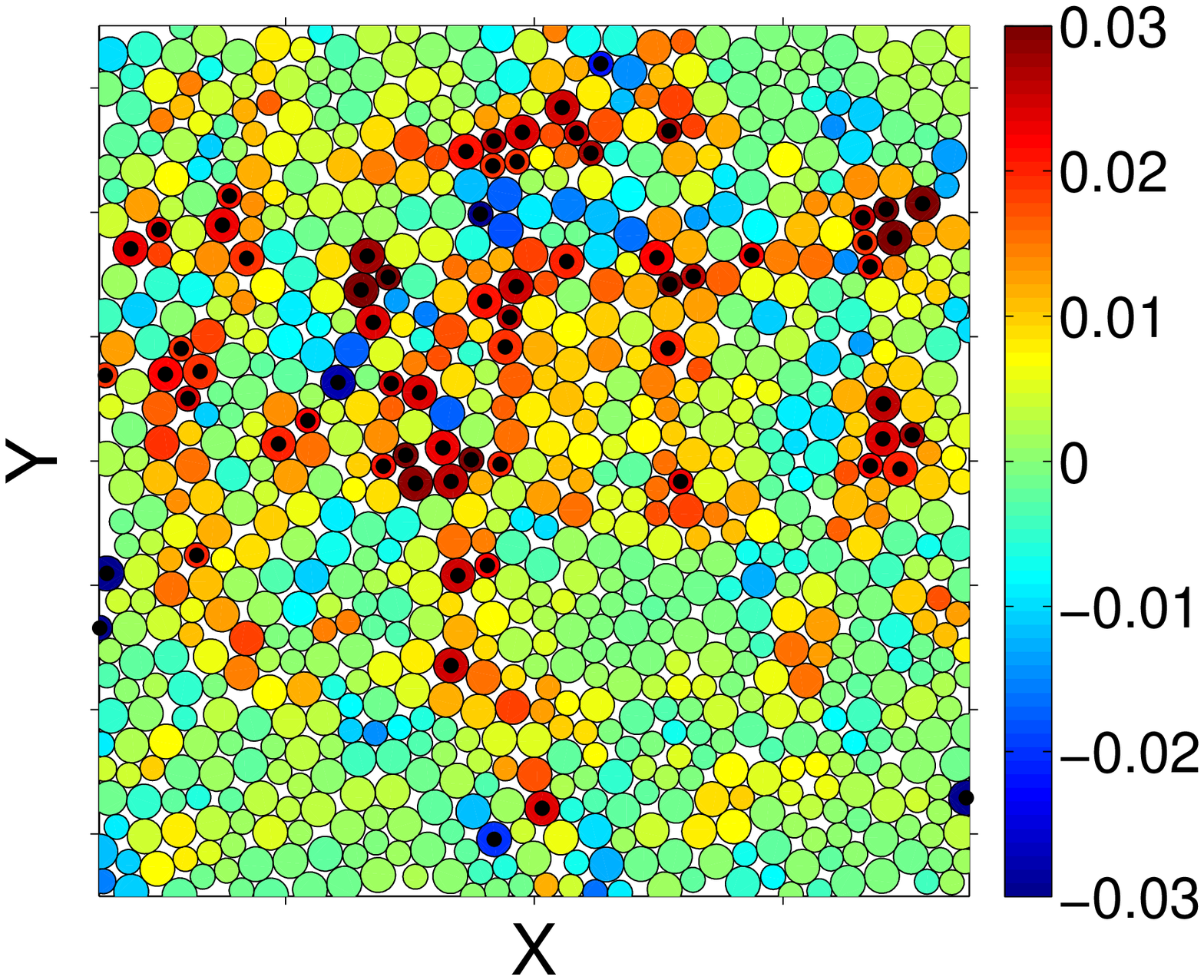}
\includegraphics[width=0.43\columnwidth,clip,viewport=228 0 690 450]{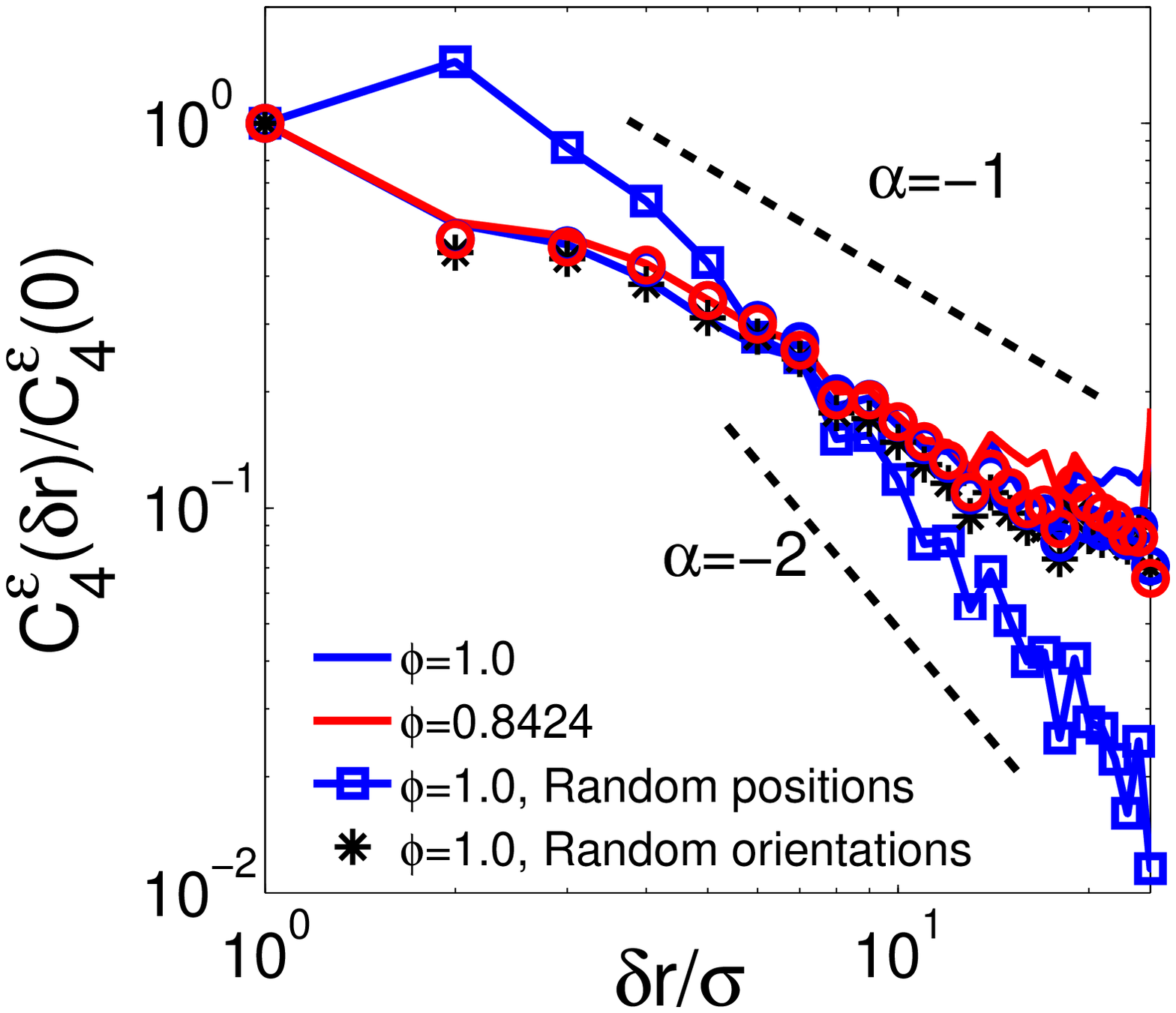}
\includegraphics[width=0.46\columnwidth,clip,viewport=203 0 687 450]{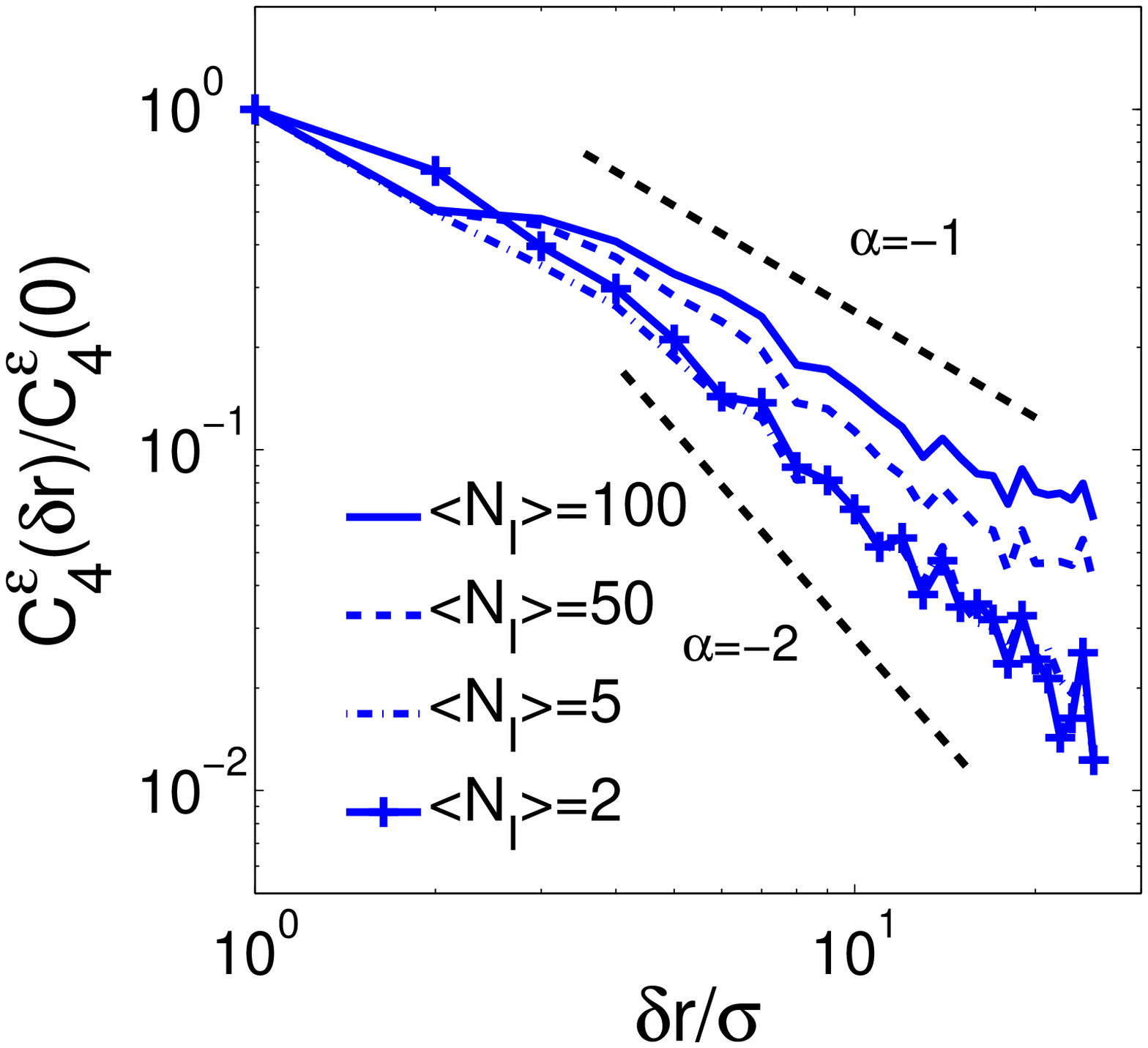}
\includegraphics[width=0.45\columnwidth,clip,viewport=200 0 700 450]{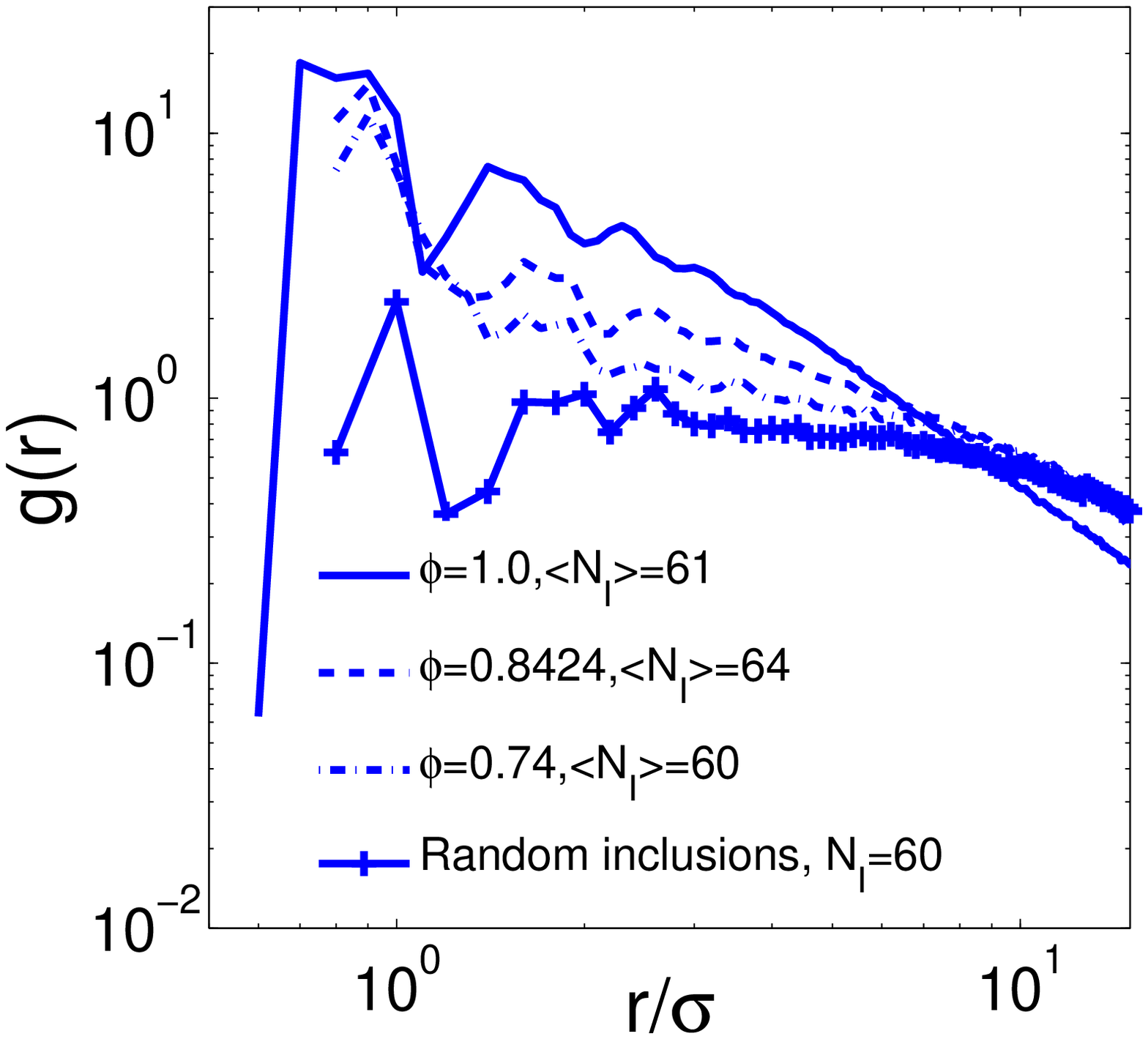}
\put(-225,140){(a)}
\put(-105,140){(b)}
\put(-225,-3){(c)}
\put(-105,-3){(d)}
\caption{Correlations of superimposed strain fields. (a) Snapshot of the instantaneous strain field and selected inclusion centers (dotted particles) at $\phi=1.0$. Particle color indicates sign and magnitude of the local shear strain ($\epsilon_{xy}$), see color bar. (b) Strain correlations as a function of $\delta r$ after quadrupolar projection, see text. Data taken from simulations at volume fractions 1.0 and 0.8424 (red line), from superimposed inclusions positioned at strain maxima (red circles), from inclusions at random positions (blue squares), and from inclusions at the right positions, but with random orientations (black asterisks). The superimposed inclusions describe the strain field correctly if they are at the right positions.
(c,d) Crossover upon variation of number of inclusions and particle volume fraction. (c) Dependence of constructed strain correlations on the number of inclusions. Taking a few top-strain inclusions leads to the trivial decay with slope -2, and does not capture the real correlations. (d) Volume fraction dependence of pair correlations of inclusions, see legend. With decreasing volume fraction, and concomitant decreasing elastic correlation length, the spatial distribution of inclusions approaches that of the randomly placed inclusions.}
\label{fig45}
\end{figure}

{\em Synthethic Strain Fields:} We now turn to synthetic strain fields that we create by superposing many Eshelby quadrupoles given by Eq.~(\ref{exy}). We show a snapshot of the instantaneous strain field for $\phi=1.0$ in Fig.~\ref{fig45}a; clearly, many locations of significant strain play a role here. We thus
constructed strain fields for 50 to 70 inclusion centers identified as particles having shear strain amplitudes larger than two times the standard deviation (dotted particles in Fig.~\ref{fig45}a). We used eq.~\ref{exy} where we took $\epsilon_0$ to be the actual strain of the particle at the inclusion center, and $a$ to be the particle diameter. We then computed the total strain field by superposition, and determined strain correlations and quadrupolar projections as before. The resulting correlation function, shown in Fig.~4b (red circles), exhibits excellent overlap with the one obtained in the simulations (red line). Hence, synthetic strain fields obtained by superposition of a sufficient number of quadrupoles can capture the nontrivial powerlaw decay of $C_4$.

Which microscopic properties of the inclusions are essential to capture these nontrivial correlations? We first tested what happened when we randomize the locactions of the quadrupole centers. We no longer chose high-strain particles, but instead selected particles randomly (again between 50 and 70) as core of inclusions; we then constructed the resulting strain field by superposition, and computed the corresponding correlation functions, see Fig.~\ref{fig45}b, blue squares. Clearly, the data does not match the nontrivial decay of the correlations and rather decays with slope -2, similar to the single inclusion. This shows that a nontrivial spatial organization of the inclusions is essential to obtain the nontrivial power-law decay of $C_4$.

We further tested the robustness of these results with respect to the orientation of the inclusions. Typically, inclusions align their directions along the principal directions of the system to minimize their interaction energy, as seen in Fig.~\ref{fig2}a. At finite shear rate, however, we may expect some randomness in the orientation of the inclusions. We therefore added a random offset angle $\theta_0$, uniformly distributed between $-\pi/4$ and $\pi/4$ in Eq.~\ref{exy}. The resulting projected strain correlation (asterisks in Fig.~\ref{fig45}(b)) shows nice overlap with the power-law data of the real inclusions as long as the angle variation is not too large, indicating that with respect to the orientations, the scaling of correlations is robust.

We also investigated variations in the number of inclusions. Clearly, using one inclusion does not work, but a sufficient number of inclusions robustly pushes the powerlaw decay away from a slope -2, as illustrated in Fig.~4c. Hence, a broad population of inclusions is needed to capture the full nontrivial behavior. Similarly, when we narrow strain amplitudes $\epsilon_0$ to one fixed value for all inclusions, the resulting correlations weakly deviate from the -1 slope, although the effect is not as strong as for randomizing the locations of the quadrupoles.  We conclude that the inclusions are strongly correlated in their position, with less strong correlations in their strength and orientation. These correlations are directly reflected in the spatial organization of the quadrupoles. To show this, we have computed pair correlations of the inclusion centers~\cite{footnote}; at $\phi = 1$, we observe much stronger short-range correlation than for the random arrangement, as shown in Fig.~\ref{fig45}d.

How does this picture change below $\phi_c$? Below jamming, elastic correlations become increasingly short ranged, and this should affect the organization of inclusions. This is indeed what we find when we investigate the inclusion pair correlations as a function of volume fraction across jamming, see Fig.~\ref{fig45}d. As the density decreases, the pair correlation function becomes increasingly flat, reflecting the diminishing spatial organization of the inclusions, until at $\phi=0.74$, the pair correlation function approaches the randomly placed inclusions (while still being distinct from it), suggesting only small remaining interactions, in agreement with the elastic correlation length shown in Fig.~\ref{fig2}.

{\em Conclusions: }
In conclusion, we have investigated the nature of the flow of foams at finite but small shear rate in a numerical model. When the packing fraction is well above jamming ($\phi>\phi_c$), isolated plastic events dominate the deformation. These plastic events self-organize due to their quadrupolar elastic interactions. Surprisingly, these elastic interactions remain robust and long-range down to the jamming transition, where the elastic modulus of the quiescent packing vanishes, and the displacement field becomes delocalized. Thus, down to and even below the jamming density, the spatial organization of flowing foams is dominated by long-range elastic interactions. This is confirmed by modeling of the strain field correlations using the well-known Eshelby elastic response of a single shear distortion. We showed that the nontrivial flow field leads to the superposition of strain fields at strongly correlated locations. These results shed new light on the nature of flow across the jamming transition: over wide ranges of density from the liquid via the marginal solid to the deeply quenched solid state, complex fluid flows arise from the superposition of elastic fields of interacting Eshelby inclusions, with interaction ranges only becoming finite below the jamming transition.


\begin{thebibliography}{99}
\bibitem{footnote}
The precise center $r_i$ of inclusion $i$ is identified as the particle exhibiting the maximum strain magnitude among its Delaunay neighbors. This ensures that no more than one of the neighboring particles can be the core of an inclusion. Pair correlations are then computed according to $g(r) = \frac{A}{\pi r^2N^2}\left<\sum_i\sum_{j\ne i}\delta(r-r_{ij})\right>$, where $r_{ij}= |r_i - r_j|$, $N$ is the total number of inclusions, and $A$ is a normalization constant.
\end{thebibliography}
\end{document}